\documentclass[12pt]{article}

\usepackage{natbib}
\usepackage[table]{xcolor}
\usepackage{graphicx}
\usepackage{subcaption}
\usepackage{amsmath}
\usepackage{dsfont}
\usepackage{placeins}
\usepackage{multirow}
\usepackage{hyperref}
\usepackage{enumerate}

\usepackage{algpseudocode}
\usepackage{amsthm}
\usepackage{hyperref}
\usepackage{mathrsfs}
\usepackage{amsfonts}
\usepackage{bbding}
\usepackage{listings}
\usepackage{appendix}
\usepackage{fancyvrb}
\usepackage{cancel}
\usepackage{algorithm}

\newcounter{probnum}
\allowdisplaybreaks

\definecolor{tabblue}{rgb}{.870588,.905882,.94902}
\definecolor{gray}{rgb}{0.7,0.7,0.7}
\definecolor{black}{rgb}{0,0,0}
\definecolor{white}{rgb}{1,1,1}
\definecolor{blue}{rgb}{0.0,0.0,1}

\definecolor{green}{rgb}{0,0.5,0}

\definecolor{yellow}{rgb}{1,0.549,0}

\definecolor{red}{rgb}{0.6,0.0,0.0}
\newcommand{\inred}[1]{\color{red}#1\color{black}}
\definecolor{darkred}{rgb}{0.9,0.4,0}

\definecolor{purple}{rgb}{0.58,0,0.827}

\definecolor{backgcode}{rgb}{0.97,0.97,0.8}
\definecolor{Brown}{cmyk}{0,0.81,1,0.60}
\definecolor{OliveGreen}{cmyk}{0.64,0,0.95,0.40}
\definecolor{CadetBlue}{cmyk}{0.62,0.57,0.23,0}

\newcommand{\qu}[1]{``{#1}''}
\newcommand{\bv}[1]{\boldsymbol{#1}}

\newcommand{\brho}{\boldsymbol{\rho}}

\newcommand{\bSigmaw}{\bv{\Sigma}_{\W}}

\newcommand{\bSigmawB}{\bv{\Sigma}_{\W_B}}

\newcommand{\bSigmaZ}{\bv{\rho}}

\newcommand{\muvec}{\bv{\mu}}

\newcommand{\Z}{\bv{Z}}

\newcommand{\I}{\bv{I}}
\newcommand{\J}{\bv{J}}
\newcommand{\Y}{\bv{Y}}
\newcommand{\W}{\bv{W}}

\newcommand{\x}{\bv{x}}

\newcommand{\w}{\bv{w}}

\newcommand{\tauhat}{\hat{\tau}}

\newcommand{\onevec}{\bv{1}}
\newcommand{\zerovec}{\bv{0}}

\newcommand{\y}{\bv{y}}

\renewcommand{\v}{\bv{v}}
\renewcommand{\muvec}{\bv{\mu}}

\newcommand{\z}{\bv{z}}

\newcommand{\bbeta}{\bv{\beta}}

\newcommand{\reals}{\mathbb{R}}

\newcommand{\beqn}{\vspace{-0.25cm}\begin{eqnarray*}}
\newcommand{\eeqn}{\end{eqnarray*}}
\newcommand{\bneqn}{\vspace{-0.25cm}\begin{eqnarray}}
\newcommand{\eneqn}{\end{eqnarray}}
\newcommand{\benum}{\begin{enumerate}}
\newcommand{\eenum}{\end{enumerate}}

\newcommand{\parens}[1]{\left(#1\right)}

\newcommand{\bracks}[1]{\left[#1\right]}
\newcommand{\braces}[1]{\left\{#1\right\}}

\newcommand{\norm}[1]{\left|\left|#1\right|\right|}
\newcommand{\normsq}[1]{\norm{#1}^2}

\newcommand{\inverse}[1]{\parens{#1}^{-1}}

\newcommand{\expe}[1]{\mathbb{E}\bracks{#1}}

\newcommand{\expesub}[2]{\mathbb{E}_{\,#1}\bracks{#2}}
\newcommand{\cexpesub}[3]{\expesub{#1}{#2\,|\,#3}}

\newcommand{\var}[1]{\mathbb{V}\text{ar}\bracks{#1}}

\newcommand{\varsub}[2]{\mathbb{V}\text{ar}_{#1}\bracks{#2}}
\newcommand{\cvarsub}[3]{\mathbb{V}\text{ar}_{#1}\bracks{#2\,|\,#3}}

\newcommand{\msesub}[2]{\mathbb{M}\text{SE}_{#1}\bracks{#2}}

\newcommand{\oneover}[1]{\frac{1}{#1}}

\newcommand{\bernoulli}[1]{\mathrm{Bernoulli}\parens{#1}}
\newcommand{\betanot}[2]{\mathrm{Beta}\parens{#1,\,#2}}

\newcommand{\normnot}[2]{\mathcal{N}\parens{#1,\,#2}}

\newcommand{\poisson}[1]{\mathrm{Poisson}\parens{#1}}

\newcommand{\weibullnot}[2]{\mathrm{Weibull}\parens{#1,\,#2}}

\newcommand{\onevecall}{\onevec_{2n}}
\newcommand{\onevecallT}{\onevecall^\top}

\newcommand{\msetext}{\text{MSE}}

\usepackage{fancyhdr}
\usepackage{abstract}
\usepackage[auth-sc,affil-sl]{authblk}
\usepackage[margin=1.0in]{geometry}
\usepackage{amssymb}
\usepackage{algpseudocode}

\newtheorem{theorem}{Theorem}[section]

\newtheorem{proposition}{Proposition}[theorem]
\newtheorem{corollary}{Corollary}[theorem]

\newtheorem{assumption}{Assumption}[section]

\newcommand{\tr}[1]{\text{tr}\bracks{#1}}

\newcommand{\ourtitle}{The Pairwise Matching Design is Optimal under Extreme Noise and Extreme Assignments}
\title{\ourtitle}

\author[1]{Adam Kapelner\thanks{Electronic address: \texttt{kapelner@qc.cuny.edu}; Principal Corresponding author}}
\author[2]{Abba M. Krieger\thanks{Electronic address: \texttt{krieger@wharton.upenn.edu}}}
\author[3]{David Azriel\thanks{Electronic address: \texttt{davidazr@technion.ac.il}}}

\affil[1]{\small Department of Mathematics, Queens College, CUNY, USA}
\affil[2]{\small Department of Statistics, The Wharton School of the University of Pennsylvania, USA}
\affil[3]{\small Faculty of Data and Decision Sciences, The Technion, Haifa, Israel}

\begin{document}
\maketitle

\begin{abstract}
We consider the general performance of the difference-in-means estimator in an equally-allocated two-arm randomized experiment under common experimental endpoints such as continuous (regression), incidence, proportion, count and uncensored survival. We consider two sources of randomness: the subject-specific assignments and the contribution of unobserved subject-specific measurements. We then examine mean squared error (MSE) performance under a new, more realistic \emph{simultaneous tail criterion}. We prove that the pairwise matching design of Greevy et al. (2004) asymptotically performs best under the simultaneous tail criterion when compared to other blocking designs for all response types. We also prove that the optimal design must be less random than complete randomization and more random than any deterministic, optimized allocation. Theoretical results are supported by simulations.
\end{abstract}


\noindent%
{\it Keywords:}  experimental design, optimal design, pairwise matching, blocking, incidence endpoint, count endpoint
\vfill

\section{Background}\label{sec:intro}

We consider a classic problem: a randomized experiment with $2n$ \emph{subjects} (\emph{individuals}, \emph{participants} or \emph{units}). The experiment has two \emph{arms} (\emph{treatments}, \emph{manipulations} or \emph{groups}), which we will call treatment and control denoted $T$ and $C$. The goal herein is to infer the causal sample average treatment effect and thus, we randomize subjects thereby assigning them into the two groups. 

Formally defined, the \emph{randomization} (\emph{allocation} or an \emph{assignment}) is a vector $\w := \bracks{w_1, \ldots, w_{2n}}^\top$ whose entries indicate whether the subject received $T$ (coded numerically as +1) or $C$ (coded numerically as -1) and thus this $\w \in \braces{-1,+1}^{2n}$. The only liberty the experimenter has is to choose the entries of $\w$. When its entries are allocated randomly, the process that results in the values of the $2n$ subjects is termed an experimental \emph{design} (\emph{strategy}, \emph{algorithm}, \emph{method} or \emph{procedure}.) Experimental design may be viewed as a random draw (most often with equal probability) from a set of $\w$'s; this process we denote as random variable $\W$.

After the experiment concludes, we measure one \emph{outcome} (\emph{response} or \emph{endpoint}) of interest for the $2n$ subjects denoted $\y = \bracks{y_1, \ldots, y_{2n}}^\top$. We consider the $y_i$ values to be one of five response types: continuous, incidence, count, proportion or uncensored survival. We then use the response $\y$ along with the assignment $\w$ to perform inference for the causal sample average treatment effect.

We show herein that the optimal asymptotic design for general response is one that blends pure randomization with deterministic optimization. Further, this optimal design is satisfied by pairwise matching. Before we provide more details about our contribution, we contextualize our result by providing background of both classic and modern experimental designs. In doing so, we will discuss our previous work of which the work herein is an important extension.

\subsection{Historical Designs}\label{subsec:historical_designs}

If nothing is known about the subjects before assignment, the typical \qu{gold-standard} design is the balanced complete randomization design (BCRD, \citealp[p. 1171]{Wu1981}). This design has $\binom{2n}{n}$ allocations each that satisfy the balance constraint $\sum_i w_i = 0$. Each allocation is then chosen with equal probability.

If the subjects have observed characteristics, other designs which incorporate these a priori measurements are also popular. Hereon, we assume each subject has $p$ real-valued observed subject-specific \emph{covariates} (\emph{measurements}, \emph{features} or \emph{characteristics}) collected into a row vector denoted $\x_i$ for the $i$th subject. The setting we investigate is where all $\x_i$'s are known beforehand and considered fixed. This non-sequential setting was studied by \citet{Fisher1925} when assigning treatments to agricultural plots and is still of great importance today. In fact, this setting is found in \qu{many phase I studies [that] use `banks' of healthy volunteers ... [and] ... in most cluster randomised trials, the clusters are identified before treatment is started} \citep[page 1440]{Senn2013}. The more common \emph{sequential} experimental setting where subjects arrive one-by-one and the experimenter is obligated to assign each on the spot (hence all $\x_i$'s are not seen beforehand), will be left for future work.

Fisher, at the time of his proposing randomized experiments, immediately noted that under some unlucky $\w$'s there are large differences in the distribution of observed covariates between the two arms for sample sizes common in clinical trials. The amount of covariate value heterogeneity between groups we term \emph{observed imbalance}. This observed imbalance creates estimation bias from the perspective of any single $\w$ (or estimation variance from the perspective of all $\w$'s).

Thus, sometimes the gold-standard BCRD is not employed and instead \qu{restricted} designs are used. These designs restrict the allowable $\w$'s in an effort to minimize the observed imbalance among the two arms. Restricted designs have a long literature once again starting with \citet[p. 251]{Fisher1925} who wrote \qu{it is still possible to eliminate much of the \ldots heterogeneity, and so increase the accuracy of our [estimator], by laying restrictions on the order in which the strips are arranged}. Here, he introduced the \emph{block design}, a restricted design still popular today, especially in clinical trials. Another early mitigation approach can be found in \citet[p. 366]{Student1938} who wrote that after an unlucky, highly imbalanced randomization, \qu{it would be pedantic to [run the experiment] with [an assignment] known beforehand to be likely to lead to a misleading conclusion}. His solution is for \qu{common sense [to prevail] and chance [be] invoked a second time}. In foregoing the first assignment and \emph{rerandomizing} to find a better assignment, all allocations worse than a predetermined threshold of observed imbalance are eliminated. This classic strategy has been rigorously investigated only recently \citep{Morgan2012, Li2018} in the case of equal allocation. Another idea is to allocate treatment and control among similar subjects by using a pairwise matching (PM) design \citep{Greevy2004} which  creates $n$ pairs of subjects which minimize an overall average covariate distance within pair. In each pair, one subject receives T and the other receives C with equal probability.

One may wonder about the \qu{optimal} restricted design strategy. The design $\W$ is a multivariate Bernoulli random variable that has $2^{2n} - 1$ parameters, an exponentially large number \citep[Section 2.3]{Teugels1990}. Finding the optimal design is thus tantamount to solving for exponentially many parameters using only the $2n$ observations, a hopelessly unidentifiable task. Thus, to find an \qu{optimal} design, the space of designs must be limited. One way of limiting the space of designs is to do away with any notion of randomization whatsoever and instead compute a deterministic perfect balance (PB) optimal allocation; this is a design with two allocation vectors $\w_*$ and the allocation where its subjects switch arms, $-\w_*$. An early advocate of this approach was \citet{Smith1918} and a good review of classic works advocating this approach such as \citet{Kiefer1959} is given in \citet[Chapter 3]{Steinberg1984}. A particularly edgy advocate of the PB design is \citet{Harville1975} who penned an article titled \qu{Experimental Randomization: Who Needs It?}. 

\subsection{The Randomness-Imbalance Tradeoff}\label{subsec:randomness_imbalance_tradeoff}

We have introduced thus far a continuum of designs from the most random (BCRD) and thus has the highest expected observed imbalance to the least random (one optimized, deterministic $\w$) which has the lowest expected observed imbalance. Other designs such as blocking, rerandomization and PM are thus located somewhere in the middle of these two extremes. A natural question to ask is \qu{do these two considerations (degree of randomness and degree of observed imbalance) trade off among each other}? The answer is yes; since randomization balances both observed and \textit{unobserved} covariates, the less random a design is, the more likely the unobserved covariate measurements will be imbalanced (demonstrated herein). This imbalance in unseen characteristics has the potential to wreak havoc on estimation error. Thus, the assumptions about these unobserved covariates and how they are incorporated into the experimental performance metric become critically important. 

This trade-off was first investigated in \citet{Kapelner2021} and followed up by \citet{Azriel2023}, where the latter we call herein our \qu{previous work}. For general response models, our previous work analyzed three unobserved covariate scenarios (I) nature provides the worst possible values of unobserved covariates, (II) the unobserved covariates are random noise that will be averaged over (i.e., colloquially, \qu{margined out}) and (III) the unobserved covariates are random noise but the concern is about values of these unobserved covariates that produce terrible estimator performance. Naturally, there was a threefold answer to what the optimal design would be. For continuous response, under (I), the experimenter should never sacrifice any randomness and thus BCRD is optimal; under (II), the experimenter can sacrifice all randomness and employ the deterministic design PB; and under (III), the experimenter should sacrifice some randomness and use a design in between the two extremes. But response types that are not continuous are more complicated. Under (I), the problem is intractable but when limited to the space of all block designs, then BCRD is the optimal design. Under (II), the deterministic design is still optimal but impossible to locate as it requires knowledge of the latent parameters within the response model. Under (III), the problem is again intractable but when limited to the space of all block designs, then block designs with few blocks are optimal (see therein for details).

Our work herein extends our previous work by examining (III) under a different tail criterion, one that we believe to be much more realistic. In our previous work, the tail was defined as a large quantile (e.g. 95\%, 99\%, etc) over the observed covariates of the expected squared error of the estimator over assignments. This previous tail criterion considers values of observed covariates that can lead to poor performance. However, the inner expectation over all assignments is troublesome because in any \textit{single} experiment, only one $\w$ is realized. Herein, we define the tail criterion to be the quantile over the joint distribution of the possible unobserved covariates \textit{simultaneous with} the possible design assignments and thus we name it the \emph{simultaneous tail criterion}. This distribution is the distribution over all possible experiments with the same subjects. Thus, minimizing this tail is akin to seeking designs that minimize experiments themselves that end in ruin. Our optimal design results are found to be fundamentally different than what our previous work found as we no longer have the luxury to consider the average assignment.

Locating the optimal design under our new criterion is, as expected, intractable. But, when limiting our design space to block designs as we did in our previous work, we find PM to be asymptotically optimal. We define our experimental setting formally in Section~\ref{sec:setup}, discuss criterions of performance and theoretical results in Section~\ref{sec:methods}, provide simulation evidence of our results in Section~\ref{sec:simulations} and conclude in Section~\ref{sec:discussion}.

\section{Problem Setup}\label{sec:setup}
The problem setup is identical to our previous work’s setup section. We cannot reprint it here because it will be flagged as plagiarism. Sorry for making you click on the link. We must however reprint Table 1 below because we reference it later in this work.

\begin{table}[ht]
    \centering
    \small
    \begin{tabular}{cc|cc|c}
       Response type name & $\mathcal{Y}$ & Model Name & Common $\mu_{T,i}$ and $\mu_{C,i}$ & $\Theta$ \\ \hline
       
       Continuous & $\reals$ & Linear & $\beta_0 + \bbeta^\top\x_i + \beta_T w_i$ & $\reals$\\ \hline
       
       Incidence & $\braces{0,1}$ &  Log Odds & \multirow{2}{*}{$\inverse{1 + e^{-(\beta_0 + \bbeta^\top\x_i + \beta_T w_i)}}$} & \multirow{2}{*}{$(0, 1)$} \\
       
       Proportion & $(0, 1)$ & Linear & & \\ \hline
       
       Count & $\braces{0,1, \ldots}$ & \multirow{2}{*}{Log Linear} & \multirow{2}{*}{$e^{(\beta_0 + \bbeta^\top\x_i + \beta_T w_i)}$} & \multirow{2}{*}{$(0, \infty)$}\\
       
       Survival & $(0, \infty)$ &  &\\ \hline
       
    \end{tabular}
       
       
       
       
       
       
    \caption{Response types considered in this work. For the example mean response model, the $\beta$'s are unknown parameters to be estimated of which $\beta_T$ is most often of paramount interest to the experimenter.}
    \label{tab:simple_models}
\end{table}

\section{Analyzing Designs}\label{sec:methods}

Typically, the mean squared error (MSE) is the metric used to measure estimator performance. Trivial computation shows our estimator of Equation~\ref{eq:estimator} is unbiased under Assumption~\ref{ass:equal_chance_T}. Thus, the expectation of $(\tauhat - \tau)^2$ over all assignments is its variance,

\bneqn\label{eq:mse}
\msesub{\W}{\tauhat} = \varsub{\W}{\tauhat} = \oneover{4n^2} (\muvec_T + \z_T + \muvec_C + \z_C)^\top \bSigmaw (\muvec_T + \z_T + \muvec_C + \z_C)
\eneqn

\noindent where the defining matrix $\bSigmaw$ denotes $\var{\W}$, the $2n \times 2n$ variance-covariance matrix of the design.

It is impossible to directly employ the MSE as the criterion in which we compare different designs since the quantities $\muvec_T, \z_T, \muvec_C, \z_C$ are not observed. Thus, we need other criterions to prove results. We now review briefly the criterions and results from our previous work. 

\subsection{Worst Case MSE Criterion}

We previously considered the criterion of the \emph{worst case MSE}, 

\beqn
\sup_{\muvec_T, \z_T, \muvec_C, \z_C} \msesub{\W}{\tauhat}
\eeqn

\noindent i.e., the MSE for the worst possible potential outcome values. After limiting the space of potential outcome values to avoid trivial infinities, we showed that the optimal design under this criterion is BCRD for continuous response (previous work's Theorem 3.1). For non-continuous response we showed the optimal design in intractable. But when considering block designs (which are defined formally in Section~\ref{subsec:pm_optimal}), the optimal design is BCRD (previous work's Theorem 3.2). We do not believe there are many more \emph{worst case MSE} performance results available under the pure randomization model.

\subsection{Mean MSE Criterion}

Our previous work then considered relaxations to the randomization model allowing for randomness in the values of $\z_T, \z_C$. We will do the same herein. We formally let $\Z_T$ and $\Z_C$ be the vector random variables that generate the $\z_T$ and $\z_C$ vectors respectively. We assume

\begin{assumption}\label{ass:z_expectation_zero}
$\expe{\Z_T} = \expe{\Z_C} = \zerovec_{2n}$.
\end{assumption}

\begin{assumption}\label{ass:z_independent}
The components in $\Z_T$ are independent, the components of $\Z_C$ are independent and the components of the former and latter are mutually independent.
\end{assumption}

Although the distributional form of both $\Z_T$ and $\Z_C$ are implicitly constrained by the response type specified by $\mathcal{Y}$, we do not make any more formal distributional assumptions beyond vague moment assumptions (see Section~\ref{subsec:pm_optimal}). After assuming this new source of randomness, our previous work then considered the \qu{mean MSE} criterion. Section~\ref{app:expe_mse} of the Supplementary Material derives the following:

\bneqn\label{eq:expe_mse}
\expesub{\Z_T, \Z_C}{\msesub{\W}{\tauhat}} = \oneover{4n^2}
  \muvec^\top \bSigmaw \muvec + 
  c_Z
\eneqn

\noindent where $\muvec := \muvec_T + \muvec_C$ and $c_Z$ is a constant with respect to the design. This criterion reflects the average MSE when averaging over all errors and misspecifications, $\z_T$ and $\z_C$.  This is a popular criterion in the literature to evaluate designs. Under this criterion, the optimal design is intractable (since $\muvec_T$ and $\muvec_C$ are unknown) except in the case of $p=1$ and the response is continuous where the optimal design is the deterministic PB. But when considering block designs (which are defined formally in Section~\ref{subsec:pm_optimal}), our previous worked showed the optimal design is PM as long as the experimenter can order the subjects by the value of $\muvec$ (previous work's Theorem 3.3). However, this presumed ordering may not be realistic since the response function and its parameters are unknown (columns 3 and 4 of Table~\ref{tab:simple_models}). Put another way, in experimental practice, only the values of the $\x_i$'s are available and knowledge of their values does not identify the order of $\muvec$.

The \emph{worst case MSE} design only considers the most improbable tail event and thus is too conservative. But on the other hand, the \emph{mean MSE} design is either intractable or might lead to idiosyncratic $\z_T$'s and $\z_C$'s that introduce large mean squared error with non-negligible probability. 

\subsection{Quantile MSE Criterion}

Thus, we wished to analyze a criterion that balances these two extremes; one such criterion is a tail event MSE. Our previous work defined the \qu{single quantile tail criterion} which measures this tail event, $\text{Quantile}_{\Z_T, \Z_C}\bracks{\msesub{\W}{\tauhat}, q}$. The optimal design is also intractable under this criterion. However, we derived the asymptotic optimal performance (previous work's Theorem 3.4) and then showed that the optimal design must be less random than BCRD and more random than PB thus harmonizing these two extremes (previous work's Theorem 3.5). Then, when considering block designs, the block design with few blocks relative to $n$ achieved the asymptotic optimal performance (previous work's Theorem 3.6).

However, we believe that this tail criterion is unrealistic; it averages over every possible assignment vector $\w$ and only thereafter considers unlucky $\z_T,\z_C$ vectors. But in the real world, any experiment has only one $\w$. This work seeks to introduce and improved quantile MSE criterion to address this concern which we turn to now.

\subsection{Simultaneous Quantile MSE Criterion}

We proffer the \qu{simultaneous quantile tail criterion} which considers potentially ruinous joint $\braces{\w, \z_T, \z_C}$ realizations,

\bneqn\label{eq:actual_quantile_criterion}
\text{Quantile}_{\W, \Z_T, \Z_C}\bracks{(\tauhat - \tau)^2, q} 
= \expesub{\W, \Z_T, \Z_C}{(\tauhat - \tau)^2} + c_{\W, q} \sqrt{\varsub{\W, \Z_T, \Z_C}{(\tauhat - \tau)^2}},
\eneqn

\noindent\noindent where $\W$ and $q$ determine the constant $c_{\W, q}$. Although the design also affects the constant $c_{\W, q}$, our simulation results of the next section (and in our previous work) indicate that $c_{\W, q}$ remains more or less constant with respect to a certain quantile considering a wide variety of designs and settings. We henceforth consider the \qu{approximate simultaneous quantile tail criterion},

\bneqn\label{eq:approx_quantile_criterion}
Q_q &:=& \expesub{\W, \Z_T, \Z_C}{(\tauhat - \tau)^2} + c_{q} \sqrt{\varsub{\W, \Z_T, \Z_C}{(\tauhat - \tau)^2}}
\eneqn

\noindent which treats $c_{\W, q}$ as a fixed value based on $q$ denoted $c_q$. We demonstrate later via simulation that employing the standard normal quantile $c_q := z_q$ provides near-exact quantiles as we conjecture there is a central limit theorem on $(\tauhat - \tau)^2$ over realizations of $\W, \Z_T, \Z_C$. We made the analogous approximation when defining the previous work's quantile MSE criterion.

This criterion allows us to incorporate the effect of both the mean and variance with respect to the unobserved covariates and allows us to compare designs asymptotically. We discuss other criterions that can capture both considerations in Section~\ref{sec:discussion}. 

The expectation term in our criterion of Equation~\ref{eq:approx_quantile_criterion} is equivalent to Equation~\ref{eq:expe_mse}. The variance term in our criterion of Equation~\ref{eq:approx_quantile_criterion} is complicated as it has fourth moments in $\W$. As in the previous work, it is impossible to locate the optimal design according to criterion $Q_q$. Its expectation term is a function of the unknown $\muvec$ and its variance term is a function of unknown moments of $\Z_T, \Z_C$ among other unknown quantities.



Our main results found in the next two sections are (1) among block designs, PM has optimal performance as measured by $Q_q$ and (2) deterministic perfect balance (PB) performs worse than PM. These two results imply the optimal design exists between these two extremes (i.e., more random than PB and less random than BCRD) similar to the previous work's results.

\subsubsection{The PM design is Optimal Among Block Designs}\label{subsec:pm_optimal}

We now restrict our analysis to block designs. Assume the $2n$ observations are divided into $B$ blocks (i.e., sets of subject indices) where each is of size $n_B:=2n/B$. We assume that the first $n_B$ observations belong to the first block, the second $n_B$ observations to the second block, and so on. For each block, all possible allocations that divide the observations into two same-size subsets (corresponding to the two treatments) are equally weighted, and the allocations between two blocks are independent. It follows that for block designs with $B$ blocks the covariance matrix, $\bSigmawB$, is a block-diagonal matrix with blocks of the form $\frac{n_B}{n_B-1}\I_{n_B}+\frac{1}{n_B-1}\J_{n_B}$, where $\I_d$ is the identity matrix of dimension $d$ and $\J_d := \onevec_d \onevec_d^\top$, the matrix of all ones and let $\W_B$ denote the block design with $B$ blocks. 

By Equation~\ref{eq:approx_quantile_criterion}, the quantile criterion is minimized for any value of $c_q$ when both the expectation and the variance of $(\hat{\tau}-\tau)^2$ are minimal. Below we show that PM is asymptotically minimal among block designs for both $\varsub{\W_B, \Z_T, \Z_C}{(\tauhat - \tau)^2}$ and $\expesub{\W, \Z_T, \Z_C}{(\tauhat - \tau)^2}$. This result will imply that PM is asymptotically optimal among all block designs regardless of the value of the $c_q$. To do so, we will need further assumptions:

\begin{assumption}\label{ass:A0}
$\expe{Z_i^4}$ is bounded for all $i$,
\end{assumption}

\begin{assumption}\label{ass:A1}
$\frac{1}{2n} \sum_{i=1}^{2n} \rho_i \to \bar{\rho}$ a constant where $\rho_i := \var{Z_i}$, 
\end{assumption}

\noindent which are standard assumptions about moments being bounded. Section~\ref{app:proof_variance_bound} proves the following result.

\begin{theorem}
\label{thm:variance_bound}
Under assumptions~\ref{ass:z_expectation_zero} - \ref{ass:A1}, for all block designs including BCRD ($B=1$) and PM ($B=n$), $\lim\inf_{n \to \infty} n^2 \,\varsub{\W_B, \Z_T, \Z_C}{(\tauhat - \tau)^2} \ge  \frac{\bar{\rho}^2}{8} $.
\end{theorem}

\noindent Thus, $\bar{\rho}^2 / 8$ is a lower bound for the asymptotic variance of the squared error for block designs. To prove our flagship result, we now make one more assumption. 

\begin{assumption}\label{ass:A2}
Under the PM design, $\lim_{n \rightarrow \infty} \frac{1}{n} \sum_{i=1}^n (\mu_{2i}-\mu_{2i-1})^2 = 0$.
\end{assumption}

We show in Section \ref{app:justifying_the_matching_condition} of the Supplementary Material that Assumption \ref{ass:A2} holds for the optimal nonbipartite pairwise matching algorithm when the covariates are bounded and $\mu_i = f(\x_i)$, where  $f$ is Lipschitz continuous. This result can be extended to the case where the covariates are sampled from a distribution with light tails. 

Section~\ref{app:proof_pm_optimal} of the Supplementary Material proves the variance's lower bound is attained under PM i.e.,

\begin{theorem}
\label{thm:pm_optimal_for_variance}
Under assumptions~\ref{ass:z_expectation_zero} - \ref{ass:A2}, $\lim_{n \to \infty} n^2 \,\varsub{\W_n, \Z_T, \Z_C}{(\tauhat - \tau)^2} =  \frac{\bar{\rho}^2}{8}$.
\end{theorem}

Furthermore, in our criterion $Q_q$ (Equation~\ref{eq:expe_mse}), we have two terms: the square root of the variance term (analyzed above in Theorem~\ref{thm:pm_optimal_for_variance}) and the expectation term $\oneover{4n^2} \muvec^\top \bSigmawB \muvec$ which for PM equals to $\oneover{4n^2}\sum_{i=1}^n (\mu_{2i}-\mu_{2i-1})^2$. By Assumption \ref{ass:A2}, this term is of order $1/n$; i.e., it is of smaller order than $\sqrt{\varsub{\W_B, \Z_T, \Z_C}{(\tauhat - \tau)^2}}$. Combining this reasoning with Theorem~\ref{thm:pm_optimal_for_variance} above implies our main result,

\begin{corollary}\label{corr:pm_optimal_for_tail}
Under Assumptions~\ref{ass:z_expectation_zero} - \ref{ass:A2}, PM is the asymptotically optimal block design when performance is measured by criterion $Q_q$ for all response types under consideration and for all values of the constant $c_q$.
\end{corollary}

\noindent This result will likely also hold for many other similar criterions which are a function of the mean and variance.

\subsubsection{PM Outperforms PB and the Optimal Design is Harmony}\label{subsec:tail_PB}

In the case of equal allocation, the PB design is composed of a single $\w_*$ and its mirror $-\w_*$ that both uniquely minimize $|(\muvec_T + \muvec_C)^\top\w|$. If $\muvec_T + \muvec_C$ were known (which they are not except in the case of the linear continuous response with $p=1$, in which they are known up to a linear transformation), the expected value term in the MSE (Equation~\ref{eq:expe_mse}) is negligible; it is of order of $O(n^3 2^{-2n})$ as shown in \citet[Section 3.3]{Kallus2018}. In Section~\ref{app:var_of_PB} of the Supplementary Material we show that

\beqn
\lim_{n \to \infty} n^2 \,\varsub{\W_n, \Z_T, \Z_C}{(\tauhat - \tau)^2} =  \frac{\bar{\rho}^2}{2}.
\eeqn

\noindent This implies that the variance of the MSE under PB is approximately four times larger than under PM (cf. Theorem~\ref{thm:pm_optimal_for_variance}). Notice that under both PB and PM, the mean MSE is negligible compared to the standard error of the MSE, as the latter is of order $1/n$ for both PB and PM and the former is of smaller order.

Thus, similar to BCRD,  PB performs worse than PM. This result combined with the demonstration that PM performs better than BCRD (Corollary~\ref{corr:pm_optimal_for_tail}) implies that the optimal design according to criterion $Q$ is more random than BCRD and less random than PB (i.e., the optimal design has less observed covariate imbalance than BCRD and more observed covariate imbalance than PB).

\section{Simulations}\label{sec:simulations}

We simulate under a variety of scenarios to verify the performance of the simple estimator (Equation~\ref{eq:estimator}) under the mean criterion (Equation~\ref{eq:expe_mse}) and the $q=.95$ tail criterion (Equation~\ref{eq:approx_quantile_criterion}) comports with our main result (Corrolary~\ref{corr:pm_optimal_for_tail}).

\subsection{Comparing Block Designs}\label{subsec:sim_comparing_block_designs}

The sample size simulated is $2n = 96$ so that we have both a realistically-sized experiment and a wide variety of homogeneously-sized blocking designs, $B \in \braces{1, 2, 3, 4, 6, 8, 12, 16, 24, 48}$. We simulate under a different number of covariates $p \in \braces{1, 2, 5}$. We use all five common response types: continuous, incidence, proportion, count and uncensored survival. All response mean models are GLMs and hence contain an internal linear component $\mu_i := \beta_0 + \bbeta_1 \x_{i} + \beta_T w_i$ (see Table~\ref{tab:simple_models}). The values of the covariate coefficients are kept constant for all response types: $\beta_0 = -1$, $\bbeta = \bracks{1~-1~1~-1~1}$ and $\beta_T = 0.001$. The covariates were chosen to be highly representative in the response function and the treatment effect was chosen to be small; these settings in tandem serve to emphasize performance differences among the block designs. One set of fixed covariates are drawn for each pair of response type and number of covariates $p$. Covariate values are always drawn iid and response values are always drawn independently. Covariate distributions, response distributions and other response parameters used in this simulation are found in Table~\ref{tab:covariate_and_response_settings}. 


\begin{table}[ht]
    \centering
    \small
    \begin{tabular}{llll}
       Response & Covariate & Response & Response\\
       type name &  Distribution &  Distribution & Parameters  \\ \hline
       
       Continuous &  $U(-1, 1)$ & $\normnot{\mu_i}{\sigma^2}$ &          $\sigma = 1$\\ 
       
       Incidence &   $U(-10, 10)$ & $\bernoulli{\mu_i}$ &                   N/A  \\
       
       Proportion &  $U(-1, 1)$ & $\betanot{\phi\mu_i}{\phi(1-\mu_i)}$ & $\phi = 2$\\ 
       
       Count &       $U(-5, 5)$ & $\poisson{\mu_i}$ & N/A \\
       
       Survival &    $U(-1, 1)$ & $\weibullnot{\mu_i / \Gamma\parens{1 + 1/k}}{k}$ 
& $k = 4$\\ \hline
    \end{tabular}
    \caption{Simulation settings and parameters.}
    \label{tab:covariate_and_response_settings}
\end{table}

The tail criterion is a functional of the distribution of MSE which is a function of $\W, \Z_T, \Z_C$. To do so, within a response type, we first use the fixed $\x_i$'s to compute $\muvec_T, \muvec_C$ via the appropriate observed covariate mean function (see column 4 of Table~\ref{tab:simple_models}). Using $\muvec_T, \muvec_C$ we then draw $\y_T, \y_C$ from the appropriate response distribution in column 3 of Table~\ref{tab:covariate_and_response_settings}. (As $\z_T := \y_T - \muvec_T$ and $\z_C := \y_C - \muvec_C$, this step constitutes drawing a realization $\z_T, \z_C$). This draw of $\y_T, \y_C$ allows us to compute $\tau$ via Equation~\ref{eq:estimand}. We now draw a single single $\w$ from the appropriate block design under consideration. This allocation allows us to compute the estimate for $\tau$ via Equation~\ref{eq:estimator}. The squared error can then be computed by simply calculating $(\tauhat - \tau)^2$. 

To estimate the $q=.95$ quantile of Equation~\ref{eq:actual_quantile_criterion}, we generate $N_y := 100,000$ such realizations and record the empirical 95th percentile for each setting. Our theory is based on the approximate quantile of Equation~\ref{eq:approx_quantile_criterion}. To estimate the approximate quantile we computed the average over the $N_y$ squared errors plus 1.645 times the empirical standard error of the $N_y$ squared errors. Additionally, we employ the nonparametric bootstrap on the $N_y$ values to provide confidence intervals.

While doing simulations, we noticed that the incidence and count outcomes can have a discrete relationship between the MSE and $B$. This is due to the difficulty of obtaining $(y_{T,i}, y_{C,i})$ dis/concordant pairs. Thus, the variance of the covariates for these responses was increased to boost the variation in the number of ties (see column 2 of Table~\ref{tab:covariate_and_response_settings}).

For our theoretical result about PM to hold, the subjects (via the entries in $\muvec$) must be ordered according to Assumption~\ref{ass:A2}. For $p=1$, simply ordering the subjects in the order of the one covariate measurement satisfies this requirement (as the $\mu_i$'s are either bounded or thin-tailed in our simulations). For $p>1$, block designs were constructed as a function of the first two covariates (as the number of blocks increases exponentially in $p$, a common problem in experimental practice). We approximate an optimal block design for two covariates by first ordering the subjects by the first covariate. Then within blocks of size $2 n_B$, we sort by the second covariate. Thus, we emphasize, simulation results for $p > 1$ will be approximate. (Note that this is not the way PM is done in practice; we address that concern in a second simulation found below).

\begin{figure}[ht]
\centering\includegraphics[width=6.5in]{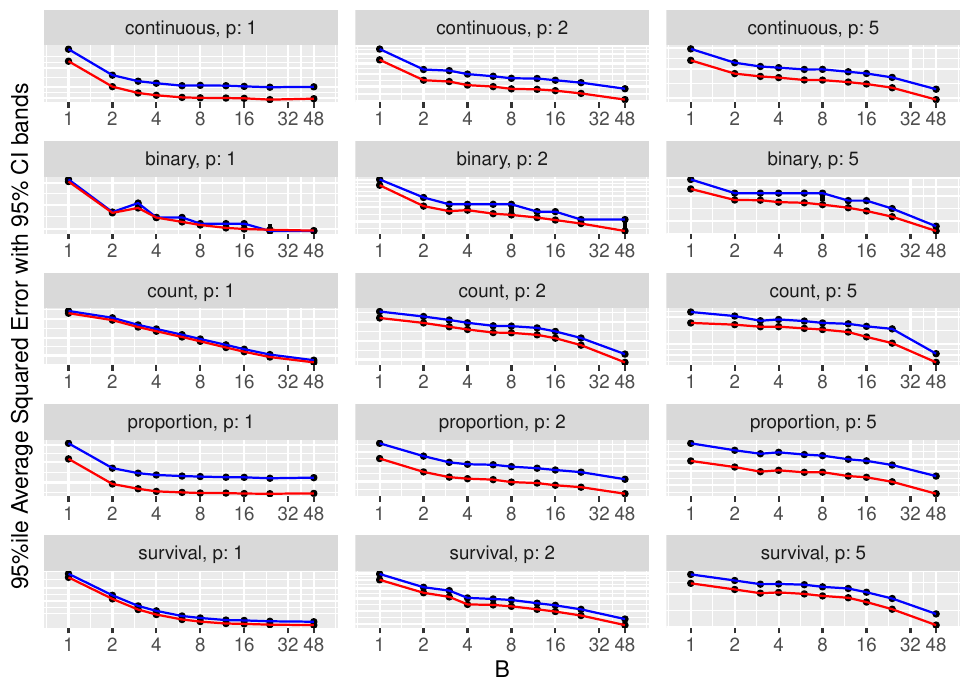} 
\caption{Simulation results for mean-centered uniform covariate realizations, $2n=96$, $p = 1,2,5$, all response types and all appropriate number of blocks. The y-axis is relative to simulation and thus its values are unshown. Blue lines are empirical 95\%ile MSE tails of and red lines are approximate 95\%ile MSE tails.}
\label{fig:all_response_by_B}
\end{figure}

The mean criterion and $q=.95$ tail criterion simulation results for $p = 1,2,5$ are presented in Figure~\ref{fig:all_response_by_B}. We can see that that PM is the best performing design for all settings even though PM is only implemented approximately using blocking on the first two covariates. 

For $p > 1$, our theorems do not exactly apply as the $\mu_i$'s are not precisely ordered to satisfy Assumption~\ref{ass:A2}, but only approximately blocked by the arrangement of two covariates. Notwithstanding, the results for $p = 2$ and $p = 5$ overall demonstrate the robustness of our theoretical results to situations of imperfect blocking and imperfect PM. Our results also demonstrate that likely the lower bound in Theorem~\ref{thm:variance_bound} is monotonically decreasing in $B$.

Further, the criterion used to prove our theorems (Equation~\ref{eq:approx_quantile_criterion}) is an approximation to the actual tail MSE (Equation~\ref{eq:actual_quantile_criterion}). Our simulations by and large demonstrate this approximation is valid. This holds true even when $\muvec$ is not ordered exactly (as in $p>1$). An additional simulation using a long-tailed covariate distribution can be found in Section~\ref{app:additional_simualtions} of the Supplementary Material. The results for the long-tailed covariate distribution are virtually identical to Figure~\ref{fig:all_response_by_B} shown here.

\subsection{Demonstrating PM is a Harmony Design}\label{subsec:sim_harmony}

In this previous simulation, PM was not implemented as one does in the practice for $p > 1$. In practice, one first calculates the proportional between-subjects Mahalanobis distance \citep[Section 2.2]{Stuart2010} via $(\x_\ell - \x_m)^\top \hat{\Sigma}_X^{-1} (\x_\ell - \x_m)$ where $\hat{\Sigma}_X^{-1}$ is the $p \times p$ sample variance covariance matrix of the $2n$ subjects' covariate vectors for all the $\binom{2n}{n}$ pairs of subjects with indices $\ell, m$. This results in a symmetric distance matrix of size $2n \times 2n$. Then, one runs the optimal nonbipartite matching algorithm that finishes in polynomial time \citep{Lu2011} which runs on the distance matrix returning the best-matched set of pairs of indices and then sorts the subjects into these pairs (the order of the pairs does not matter as $\bSigmaw$ will remain the same). We run a second simulation to both (a) check our PM performance as measured by the simultaneous $Q$ for $p > 1$ and (b) ensure PM also outperforms both PB and BCRD as a check our result of Section~\ref{subsec:tail_PB}. Thus, we compare only the BCRD, PM and PB designs. 

The second simulations' settings were the same as the first simulation except $N_y = 30,000$. For PM, we used the procedure outlined in the previous paragraph implemented in the R package \texttt{nbpMatching} \citep{Beck2016}. For $p=1$ for all response types, this procedure is identical to sorting the $\v$ vector but for $p > 1$, it can only approximately sort the $\v$ vector. For PB, recall the location of $\w_*$ is an NP-hard problem. To approximate $\w_*$, we used the greedy pair switching procedure of \citet{Krieger2019} using 10,000 starting allocation vectors and letting the best ending allocation vector of the 10,000 being considered $\w_*$. As PB is a deterministic procedure, this one $\w_*$ allocation vector was retained for all $N_y$ draws of the response values.

\begin{figure}[ht]
\centering\includegraphics[width=6.5in]{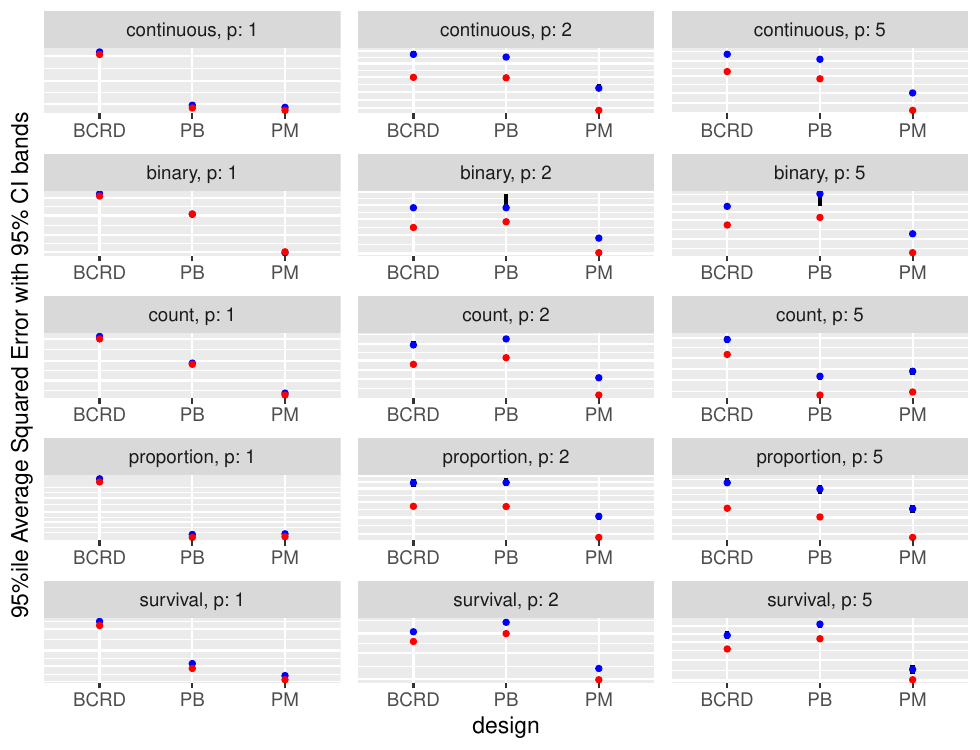} 
\caption{Simulation results for mean-centered uniform covariate realizations, $2n=96$, $p = 1,2,5$, all response types and three designs: BCRD, PB and PM. The y-axis is relative to simulation and thus its values are unshown. Blue lines are empirical 95\%ile MSE tails of and red lines are approximate 95\%ile MSE tails.}
\label{fig:all_response_pm_pb}
\end{figure}

The results of the second simulation can be found in Figure~\ref{fig:all_response_pm_pb}. Here we see that PM outperforms both PB and BCRD as measured by $Q_{.95}$ for nearly all simulations (as expected by Corollary~\ref{corr:pm_optimal_for_tail}). PM outperforming PB is expected as (a) the true $\w_*$ is unknowable from the observed covariate values alone and (b) the lack of randomness in the PB design increases the second term in the tail criterion (Equation~\ref{eq:actual_quantile_criterion}). We also see that for $p > 1$ (which our theory does not describe), the same results hold as the $\v$ vector is approximately sorted and sectioned into pairs via the nonbipartite matching procedure run on the covariate values. Also, at first glance, it is surprising that PM outperforms PB for the continuous response with $p=5$. But keep in mind that PB is computing an allocation purely on the $\x_i$'s without access to the uneven covariate weights, $\bbeta$. Given that the first covariate is weighted within the response as much as the other four covariates, making the value of $x_1$ similar within subjects pairs, PM provides superior estimation performance compared to making all five covariate averages in the two arms similar (the goal of the PB algorithm).

This empirical result also vindicates the assumptions made about the blocking design (Assumptions~\ref{ass:A0}, \ref{ass:A1} and \ref{ass:A2}) and additionally demonstrates the asymptotic results apply in the realistic experimental sample size of $n \approx 100$. 

\section{Discussion}\label{sec:discussion}

Randomized experimentation comes with the choice of \qu{how to randomize}? A default design is complete randomization independent of subjects' covariates. This comes with the risk of bias due to imbalances in subject-specific characteristics across the two arms. Can we design one perfect assignment that minimizes covariate imbalance across arms? Although this may \qu{feel} optimal, it is not (except in the case of $p=1$ and a linear response) as the covariate-response models are arbitrarily complicated. Further, this \qu{optimality} comes with great risk of the unobserved covariates creating experimental error.

Commonly employed designs are between the two extremes mentioned above; they sacrifice some randomness to get some imbalance reduction. Our previous work investigated how to harmonize these two considerations for a continuous response to choose a design. Therein we proposed a \qu{single quantile tail criterion} which averages MSE over the design and then takes the quantile of the MSE over unobserved covariate realizations. We showed this a performance metric is a trade-off of both degree of randomness and degree of observed covariate imbalance reduction. 

Herein, we propose a more realistic related criterion called the \qu{double quantile tail criterion} which takes the quantile of the MSE over both the experimental allocation and the unobserved covariates for general response types. We then prove that the asymptotically optimal block design is achieved with pairwise matching (PM) and this is the design we advise for experimenters. Although this work focuses on estimation, these blocking designs should have high power as well as many allocations can be generated which are highly independent \citep{Krieger2020b}. Our simulations were performed under small values of the average treatment effect and the results demonstrated that the design difference performance is especially sensitive to PM versus other blocking designs. Thus, taking our advice can make the difference between finding the effect or not.

There are many other avenues to explore. First, although we proved that PM is asymptotically optimal, we did not prove it is uniquely optimal as we only considered block designs. We only proved its optimality under an assumption on the differences of means within pairs. Section~\ref{app:justifying_the_matching_condition} in the Supplementary Material justifies this assumption under a suboptimal nonbipartite matching algorithm. We would like to vindicate this assumption under the real-world setting which employ Mahalanobis-distance matching in the optimal nonbipartite matching algorithm. Similar to this avenue, the results of Figure~\ref{fig:all_response_by_B} in Section~\ref{sec:simulations} and Figure~\ref{fig:exp} in Section~\ref{app:additional_simualtions} of the Supplementary Material show the performance increasing monotonically in $B$. Proving this conjecture can reveal more intuition about how PM is optimal. It can also help us find the optimal imbalance design under unequal allocation (where the number of treatments differ in the two arms). 

Another area of exploration could be understanding the tail criterion for other designs such as rerandomization and greedy pair switching. Additionally, survival is an important experimental endpoint in clinical trials but this work only considered uncensored survival measurement while the vast majority of survivals collected in real-world experiments are censored. We conjecture that our results should hold as realistic censoring mechanisms do not change the response distributions. 

Also, in \citet[Section 2.3]{Kapelner2021}, we explored the ordinary least squares (OLS) estimator for continuous response. We believe the performance of the OLS estimator in our tail criterion can be analyzed and our intuition is that PM would also emerge as optimal among block designs. However, this avenue would be difficult to explore for general response types as the analogous estimators do not have closed form (e.g. if the response is binary, the multivariate logistic regression estimates are the result of computational iteration). Further, our tail criterion is only one metric that combines mean and standard deviation of MSE. Other criterions can also be useful such as their ratio (e.g., the coefficient of variation). We can also investigate optimal designs for sequential experiments, the most common type of clinical trial.

\section*{Funding}

This research was supported by Grant No 2018112 from the United States-Israel Binational Science Foundation (BSF).

\bibliographystyle{apalike}
\bibliography{refs}

\appendix
\pagebreak

\begin{center}
   \Large{Supplementary Materials for} \\ 
   \LARGE{\qu{\ourtitle}} 
\\~\\ \large{by David Azriel, Abba Krieger and Adam Kapelner}
\end{center}

\section{Technical Proofs}

\subsection{Estimator Computation}\label{app:estimator_computation}

The contribution to  $\hat{\tau}$ from individual $i$ is $y_{T,i}/n$ when $w_i=1$ and $-y_{C,i}/n$ when $w_i=-1$. This contribution can be summarized as $((y_{T,i}-y_{C,i})+w_i(y_{T,i}+y_{C,i}))/(2n)$. When this is summed over all individuals $i$, Equation~\ref{eq:estimand} results.


\subsection{Expectation of the MSE}\label{app:expe_mse}

For this section and following sections, let $\Z := \Z_T + \Z_C$ and $\muvec := \muvec_T + \muvec_C$. 

\beqn
&=& \expesub{\Z_T, \Z_C}{\expesub{\W}{(\tauhat - \tau)^2}} \\
&=& \expesub{\Z_T, \Z_C}{\oneover{4} 
\parens{
    \frac{\muvec_T + \Z_T}{n} + \frac{\muvec_C + \Z_C}{n}
    }^\top \bSigmaw \parens{
    \frac{\muvec_T + \Z_T}{n} + \frac{\muvec_C + \Z_C}{n}
    }
}\\
&=& \oneover{4n^2} \expesub{\Z}{ 
\parens{
    \muvec + \Z
    }^\top \bSigmaw \parens{
    \muvec + \Z
    }
}\\
&=& \oneover{4n^2} \expesub{\Z}{ 
  \muvec^\top \bSigmaw \muvec +
  2\muvec^\top \bSigmaw \Z +
  \Z^\top \bSigmaw \Z
}\\
&=& \oneover{4n^2} \parens{\muvec^\top \bSigmaw \muvec + \expesub{\Z}{
  \Z^\top \bSigmaw \Z
}}\\
&=& \oneover{4n^2} \parens{\muvec^\top \bSigmaw \muvec + \tr{\bSigmaw\bSigmaZ}}\\
\eeqn

\noindent where $\bSigmaZ$ denotes the diagonal variance-covariance matrix of $\Z$. To compute the trace, we need to consider only the diagonal entries of $\bSigmaw \bSigmaZ$. The $i$th diagonal entry of this product is ${\bv{\Sigma}_{W}}_{i\, \cdot} {\bv{\Sigma}_{Z}}_{\cdot\, i} = {\bv{\Sigma}_{W}}_{i\, i} {\bv{\Sigma}_{Z}}_{i\, i}$ due to $\bSigmaZ$ being diagonal. From Assumption~\ref{ass:equal_chance_T}, we know that all the diagonal entries of $\bSigmaw$ are 1 and thus $\tr{\bSigmaw \bSigmaZ} = \tr{\bSigmaZ}$. Upon substitution, the result is

\beqn
\expesub{\W, \Z_T, \Z_C}{(\tauhat - \tau)^2} &=& \oneover{4n^2} \parens{\muvec^\top \bSigmaw \muvec + \tr{\bSigmaw\bSigmaZ}} 
+ \underbrace{\oneover{4n^2} \tr{\bSigmaZ}}.
\eeqn

\noindent where the underbraced term is design-independent and denoted as $c_Z$.

\subsection{Proof of Theorem \ref{thm:variance_bound}}\label{app:proof_variance_bound}

By Equation~\ref{eq:estimator}, the MSE can be expressed as
\[
\msetext := (\tauhat - \tau)^2 = \oneover{4n^2} \parens{ \W^\top \parens{
            \Y_{T} + 
            \Y_{C}
    }}^2
\]

\noindent where this notation depresses its dependence on $\W, \Z_T, \Z_C$. The variance of the MSE is

\begin{equation}\label{eq:MSE_decomp}
\varsub{\Z,\W}{\msetext}=\varsub{\Z}{\cexpesub{\W}{\msetext}{\Z}}+\expesub{\Z}{\cvarsub{\W}{\msetext}{\Z}}
\end{equation}

\noindent where $\Z$ denotes the collection of $\Z_T, \Z_C$. Consider the variance decomposition in Equation~\ref{eq:MSE_decomp}. 
It follows that 
\[
\varsub{\Z,\W_B}{\msetext} \ge \expesub{\Z}{\cvarsub{\W_B}{\msetext}{\Z}}.
\]
Below the latter term is analyzed.

Recall that $\v:=\muvec+\Z$.
Consider first the case of $B=n$, i.e., PM.  
We have that
\[
\v^T \W_n = U_1 ( v_2 - v_1) + U_2 (v_4 - v_3) + \cdots + U_n (v_{2n} - v_{2n-1}),
\]
where $U_1,\ldots,U_n$ are iid with $P(U=1)=P(U=-1)=1/2$.  Since $U_i^2=1$, 
\[
4 n^2 \msetext =(\v^T \W_n)^2 = \sum_{i=1}^n (v_{2i}-v_{2i-1})^2 + 2 \sum_{1 \le i < j \le n} U_i U_j (v_{2i}-v_{2i-1})(v_{2j}-v_{2j-1})  
\]
The variance of $U_i U_j$ for $i \ne j$ is 1 and the covariance between $U_i U_j$ and $U_{i'} U_{j'}$ is 0 if $(i,j) \ne (i',j')$, $i<j,i'<j'$; therefore,
\begin{equation}\label{eq:var_Wn}
\cvarsub{\W_n}{ \msetext }{\Z} = \frac{1}{16 n^4} \sum_{1 \le i < j \le n} (v_{2i}-v_{2i-1})^2(v_{2j}-v_{2j-1})^2.     
\end{equation}
Thus,

\begin{multline} \label{eq:computation1}
\expesub{\Z}{\cvarsub{\W_n}{\msetext}{\Z}}\\
=\frac{1}{16n^4} \sum_{1 \le i < j \le n} [ (\mu_{2i}-\mu_{2i-1}  )^2(\mu_{2j}-\mu_{2j-1})^2 +  (\mu_{2i}-\mu_{2i-1}  )^2 (\rho_{2j}+\rho_{2j-1}  )  + (\mu_{2j}-\mu_{2j-1}  )^2 (\rho_{2i}+\rho_{2i-1}  )]\\
+\frac{1}{16n^4} \sum_{1 \le i < j \le n} (\rho_{2i}+\rho_{2i-1}  )(\rho_{2j}+\rho_{2j-1}) .    
\end{multline}

Hence,

\begin{multline}\label{eq:LB1}
\expesub{\Z}{\cvarsub{\W_n}{\msetext}{\Z}} \ge \frac{1}{16 n^4} \sum_{1 \le i < j \le n} (\rho_{2i}+\rho_{2i-1}  )(\rho_{2j}+\rho_{2j-1})\\
= \frac{1}{ 32 n^4} \sum_{ i,j \in \{1,\ldots,n\}} (\rho_{2i}+\rho_{2i-1}  )(\rho_{2j}+\rho_{2j-1})+O(1/n^3) =  \frac{1}{ 32 n^4} (\onevecallT \brho)^2 + O(1/n^3) = \frac{\bar{\rho}^2}{8 n ^2} + o(1/n^2).
\end{multline}

Consider now a block design with $B$ blocks and recall that $\W_B$ is a random allocation from this design. The vector $\W_B$ can be thought of as a mixture over pairwise designs in the following way. Let $2n_B$ be the number of subjects in each block. Let $P_b$ be a random partition of the $2n_B$ subjects into $n_B$ pairs for block $b$. A random partition is then $P_J := P_1 \times P_2 \times \ldots \times P_B$. Each random partition produces $n$ pairwise matches. Then

\begin{align*}
 \cvarsub{\W_B}{\msetext}{\Z} &= \cexpesub{P_J}{\cvarsub{\W_B}{\msetext}{\Z,P_J} }{
\Z} +\cvarsub{P_J}{\cexpesub{\W_B}{\msetext}{\Z,P_J} }{\Z}\\
&\ge \cexpesub{P_j}{\cvarsub{\W_B}{\msetext}{\Z,P_J} }{\Z}
\end{align*}
and therefore,
\[
\expesub{Z}{\cvarsub{\W_B}{\msetext}{\Z}} 
\ge \expesub{Z}{ \cexpesub{P_J}{\cvarsub{\W_B}{\msetext}{\Z,P_J} }{\Z}} = \expesub{Z,P_J} {\cvarsub{\W_B}{\msetext}{\Z,P_J}}.  
\]
Since given $P_J$, $\W_B$ is a PM design, Equation~\ref{eq:LB1} implies that for every $P_J$
\[
\cexpesub{Z} {\cvarsub{\W_B}{\msetext}{\Z,P_J}}{P_J} \ge \frac{\bar{\rho}^2}{8 n ^2} + o(1/n^2),
\]
which completes the proof of the part (a) because
\[
\expesub{P_J}{\cexpesub{Z} {\cvarsub{\W_B}{\msetext}{\Z,P_J}}{P_J}}=\expesub{Z,P_J} {\cvarsub{\W_B}{\msetext}{\Z,P_J}}.
\]

\subsection{Proof of Theorem~\ref{thm:pm_optimal_for_variance}}\label{app:proof_pm_optimal}

Consider the variance decomposition in Equation~\ref{eq:MSE_decomp}. Theorem 3.4 of our previous paper  implies that if the fourth moment of $Z_i$ is bounded (Assumption \ref{ass:A0}) then for PM

\[
\varsub{\Z}{\cexpesub{\W_n}{\msetext}{\Z}}=O(1/n^3).
\]

\noindent Hence, when multiplying by $n^2$ this term vanishes. 

Consider now $\expesub{\Z}{\cvarsub{\W_n}{\msetext}{\Z}}$. 
By Equation~\ref{eq:computation1},

\bneqn
n^2 \expesub{\Z}{\cvarsub{\W_n}{\msetext}{\Z}} 
&=& \frac{1}{16 n^2} \sum_{1 \le i < j \le n}  (\mu_{2i}-\mu_{2i-1}  )^2(\mu_{2j}-\mu_{2j-1})^2 + \label{eq:computation2a} \\
&& \frac{n-1}{8 n^2} \sum_{i=1}^n  (\mu_{2i}-\mu_{2i-1}  )^2 (\rho_{2i}+\rho_{2i-1}) +\label{eq:computation2b}\\ 
&& \frac{1}{16 n^2} \sum_{1 \le i < j \le n} (\rho_{2i}+\rho_{2i-1}  )(\rho_{2j}+\rho_{2j-1}). \label{eq:computation3}
\eneqn

\noindent The term in Equation~\ref{eq:computation2a} vanishes asymptotically because

\begin{align*}
\frac{1}{n^2} \sum_{1 \le i < j \le n}  (\mu_{2i}-\mu_{2i-1}  )^2(\mu_{2j}-\mu_{2j-1})^2
&\le \frac{2}{n^2} \sum_{i=1}^n \sum_{j=1}^n  (\mu_{2i}-\mu_{2i-1}  )^2(\mu_{2j}-\mu_{2j-1})^2\\
&= 2 \parens{\frac{1}{n} \sum_{i=1}^n  (\mu_{2i}-\mu_{2i-1}  )^2 }^2    
\end{align*}

\noindent which goes to 0 by the assumption that $\frac{1}{n} \sum_{i=1}^n (\mu_{2i}-\mu_{2i-1})^2 \to 0$. The term in Equation~\ref{eq:computation2b} vanishes because we assumed that $\frac{1}{n} \sum_{i=1}^n (\mu_{2i}-\mu_{2i-1})^2 \to 0$ and that the $\rho$'s are bounded. Thus, the dominant term is in Equation~\ref{eq:computation3}, and it is equal to

\beqn\footnotesize
\frac{1}{16n^2} \sum_{1 \le i < j \le n} (\rho_{2i}+\rho_{2i-1}  )(\rho_{2j}+\rho_{2j-1}) &=& \frac{1}{ 32n^2} \sum_{i,j \in \{1,\ldots, n\} } (\rho_{2i}+\rho_{2i-1}  )(\rho_{2j}+\rho_{2j-1}) + O\parens{n^{-1}} \\
&=& \frac{1}{ 8}\parens{\frac{1}{2n}\sum_{i=1}^{2n} \rho_i }^2 + O\parens{n^{-1}} \\
&=& \frac{\bar{\rho}^2}{8} + o(1)
\eeqn

\noindent which completes the proof. \qed

\subsection{The variance of the MSE under PB}\label{app:var_of_PB}

 Consider $\w_*$ and its mirror $-\w_*$, vectors that both uniquely minimize $|(\muvec_T + \muvec_C)^\top\w|$. The PB design is composed of the random variable $\W$ that draws $\w_*$ and $-\w_*$ each with probability 1/2. Thus,  conditional on $\Z$,
 
\[
\msetext := (\tauhat-\tau)^2 = \oneover{4n^2} \parens{ \W^\top \parens{
            \Y_{T} + 
            \Y_{C}
    }}^2
\]

\noindent is constant implying that $\cvarsub{\W}{\msetext}{\Z}=0$. The decomposition of the variance of the MSE (Equation~\ref{eq:MSE_decomp}) is thus only

\[
\varsub{\Z,\W}{\msetext}=\varsub{\Z}{\cexpesub{\W}{\msetext}{\Z}}.
\]

\noindent This expression was computed in Sections 3.3.3 and A.9  of \citet{Azriel2023} where it was shown that under  Assumptions~\ref{ass:A0}, \ref{ass:A1} and \ref{ass:A2}, 
\[
\varsub{\Z}{\cexpesub{\W}{\msetext}{\Z}}=\frac{1}{16n^4}\parens{ O(n) + 2\parens{\sum_{i=1}^{2n} \rho_i}^2} =\frac{1}{16n^4}\parens{ O(n) + 8n^2\parens{\bar{\rho}^2}}. 
\]
It follows that under PB, 

\beqn
\lim_{n \to \infty} n^2 \,\varsub{\W_n, \Z_T, \Z_C}{(\tauhat - \tau)^2} =  \frac{\bar{\rho}^2}{2}.
\eeqn

\subsection{A Justification of the Condition of Assumption~\ref{ass:A2}}\label{app:justifying_the_matching_condition}

We assume that the covariates values $x_{i,j}$ where $i = 1, \ldots, 2n$ and $j = 1,\ldots p$ live within a compact support so that $\max_{i\le n} x_{i,j} - \min_{i \le n} x_{i,j} \le B_{j}$ for all $j,n$. If the original covariate is categorical, we assume it to be dummy-coded with $p$ adjusted accordingly. As in Equation~\ref{eq:simple_components} of the text, $\mu_i := f(\x_i)$ and we now assume it is Lipschitz continuous.

The optimal nonbipartite pairwise matching algorithm creates an order of the subjects into $n$ pairs of individuals so that the quantity $\sum_{i=1}^n 
\|\x_{2i} - \x_{2i - 1}\|^2$ is minimized where $\{1,2\}$ are the indices of the first pair, $\{3,4\}$ are the indices of the second pair, etc. Usually, the Mahalanobis distance is used for the nonbipartite pairwise matching algorithm, but we can assume (by applying a linear transformation) that the covariance matrix is the identity matrix and in this case, the algorithm based on Mahalanobis distance is equivalent to obtaining a solution that minimizes $\sum_{i=1}^n \|\x_{2i} - \x_{2i - 1}\|^2$.

\begin{proposition}{}
Suppose that the covariates are bounded and $\mu_i = f(\x_i)$, where  $f$ is Lipschitz continuous, then the optimal nonbipartite pairwise matching algorithm satisfies Assumption~\ref{ass:A2}.



\begin{proof}
It is enough to show that Assumption~\ref{ass:A2} holds for a certain matching scheme since the optimal matching has the lowest value of $\sum_{i=1}^n \|\x_{2i} - \x_{2i - 1}\|^2$. 
We consider the suboptimal matching procedure below.

\begin{algorithm}
\caption*{\textbf{Algorithm } A suboptimal nonbipartite matching algorithm}\label{alg:suboptimal_matching}
~\\\textbf{Step 1}\\
Let $m := n^{1/(2p)}$, $r := n/m$, $A_{g,j} := \bracks{x_{2r(g-1) + 1,j}, x_{2rg,j}}$ for $g=1,\ldots,m$ and $j=1,\ldots,p$ where $x_{i,j}$ is redefined as the $i$th largest value in covariate $j$. Assume that $m$ and $r$ are integers for the ease of exposition. This implies there are $2r$ subjects in each of $A_{g,j}$. \\

\textbf{Step 2}\\
Based on the intervals $A_{g,j}$ we define $m^p=\sqrt{n}$ groups. We assign individuals into groups such that two individuals $i_1,i_2$ are in the same group if $x_{i_1.j}$ and $x_{i_2,j}$ are in the same interval $A_{g,j}$ for all covariates $j$. (This implies that individuals in the same group have similar values for all the $p$ covariates.)\\


\textbf{Step 3}\\
We randomly create pairs of individuals within each of the $\sqrt{n}$ groups.\\

\textbf{Step 4}\\
Collect an \qu{overflow} group corresponding to all individuals who are not matched because they belong to a group where the number of subjects is odd. Randomly pair individuals in this overflow group.
\end{algorithm}

\noindent Let $\mathcal{K}$ define the set of matched pairs $\{i_k, i_k'\}$ according to the above Algorithm. Due to the Lipschitz assumption, it is sufficient to demonstrate that

\beqn
\lim_{n \rightarrow \infty} \frac{1}{n} \sum_{k \in \mathcal{K}}\sum_{j=1}^p (x_{i_k.j} - x_{i_{k'},j})^2 = 0.
\eeqn

\noindent Since there are $\sqrt{n}$ groups for matching, the overflow group is at most of size $\sqrt{n}$, hence those matched pairs cannot affect the left-hand side of the above equation. 

For the matched pairs in step (3), denoted by $\mathcal{K}_3$, and for each covariate $j$,

\beqn
\sum_{k \in \mathcal{K}_3}(x_{i_k,j} - x_{i_{k'},j})^2 \leq r \sum_{g=1}^m d_{g,j}^2 \leq r \parens{\sum_{g=1}^m d_{g,j}}^2 \leq r B_j^2,
\eeqn

\noindent where $d_{g,j} := x_{2rg,j} - x_{2r(g-1) + 1,j}$, that is, $d_{g,j}$ is the maximal difference of covariate $j$ in the interval $A_{g,j}$ . Aggregating the above over all $j$ and dividing by $n$, we have

\beqn
\oneover{n} \sum_{j=1}^p \sum_{k \in \mathcal{K}_3}(x_{j,i_k} - x_{j,i_{k'}})^2 \leq \frac{r}{n} \sum_{j=1}^p B_j^2 = \oneover{m} \sum_{j=1}^p B_j^2 = \oneover{n^{1/(2p)}} \sum_{j=1}^p B_j^2,
\eeqn

\noindent which goes to zero as $n$ increases.

\end{proof}
\end{proposition}

\paragraph{Remark:} If the $\x_i$'s are iid, then Assumption~\ref{ass:A2} holds with probability 1 as long as the distribution does not have fat tails. Specifically, if the underlined distribution is multivariate normal, then the above $B_j$'s are of order $\sqrt{\log(n)}$ with probability 1 and then a similar argument applies. 

\section{Additional Simulations}\label{app:additional_simualtions}

\begin{figure}[ht]
\centering\includegraphics[width=6.5in]{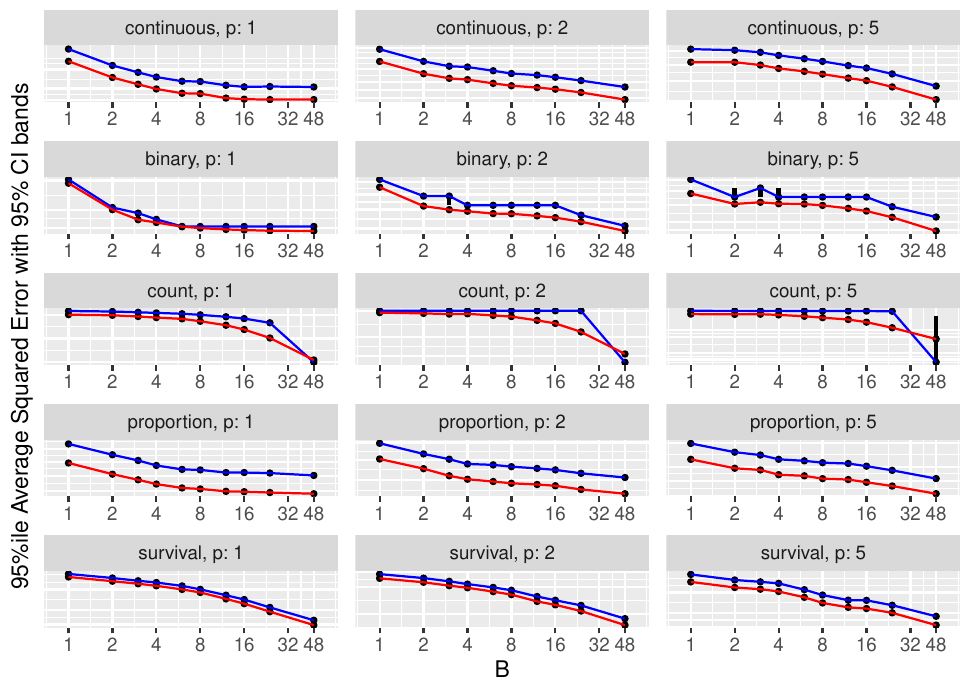} 
\caption{Simulation results for mean-centered exponential covariate realizations, $2n=96$, $p = 1,2,5$, all response types and all appropriate number of blocks. The y-axis is relative to simulation and thus its values are unshown. Blue lines are empirical 95\%ile MSE tails of and red lines are approximate 95\%ile MSE tails.}
\label{fig:exp}
\end{figure}

We run an identical simulation as in Section~\ref{sec:simulations} except we change the covariate distribution to be long-tailed (the mean-centered exponential) with parameter $\sqrt{12} / 2$ for all response types except incidence which has parameter $\sqrt{12} / 20$ and count which has parameter $\sqrt{12} / 10$. These distributions were calibrated to have the same mean (zero) and variances as the covariate distribution found in the main text (cf. Table~\ref{tab:covariate_and_response_settings}). The results are found in Figure~\ref{fig:exp} and are qualitatively similar as the simulations found in the main text. This implies that our results and asymptotics are robust to the setting of long-tailed observed subject measurements.

\end{document}